\documentclass[submission,copyright]{eptcs}
\providecommand{\event}{arXiv.org} 
\usepackage{breakurl,graphicx}             

\title{ProofPeer: Collaborative Theorem Proving \\ {\Large \emph{A Position Paper}}}
\author{Steven Obua
\qquad
Jacques Fleuriot
\qquad
Phil Scott
\qquad
David Aspinall\\
\institute{School of Informatics\\University of Edinburgh\\Scotland, United Kingdom}
}
\def\titlerunning{ProofPeer: Collaborative Theorem Proving}
\def\authorrunning{S. Obua, J. Fleuriot, P. Scott \& D. Aspinall}
\begin{document}
\maketitle

\begin{center}
\fbox{\includegraphics[width=10cm]{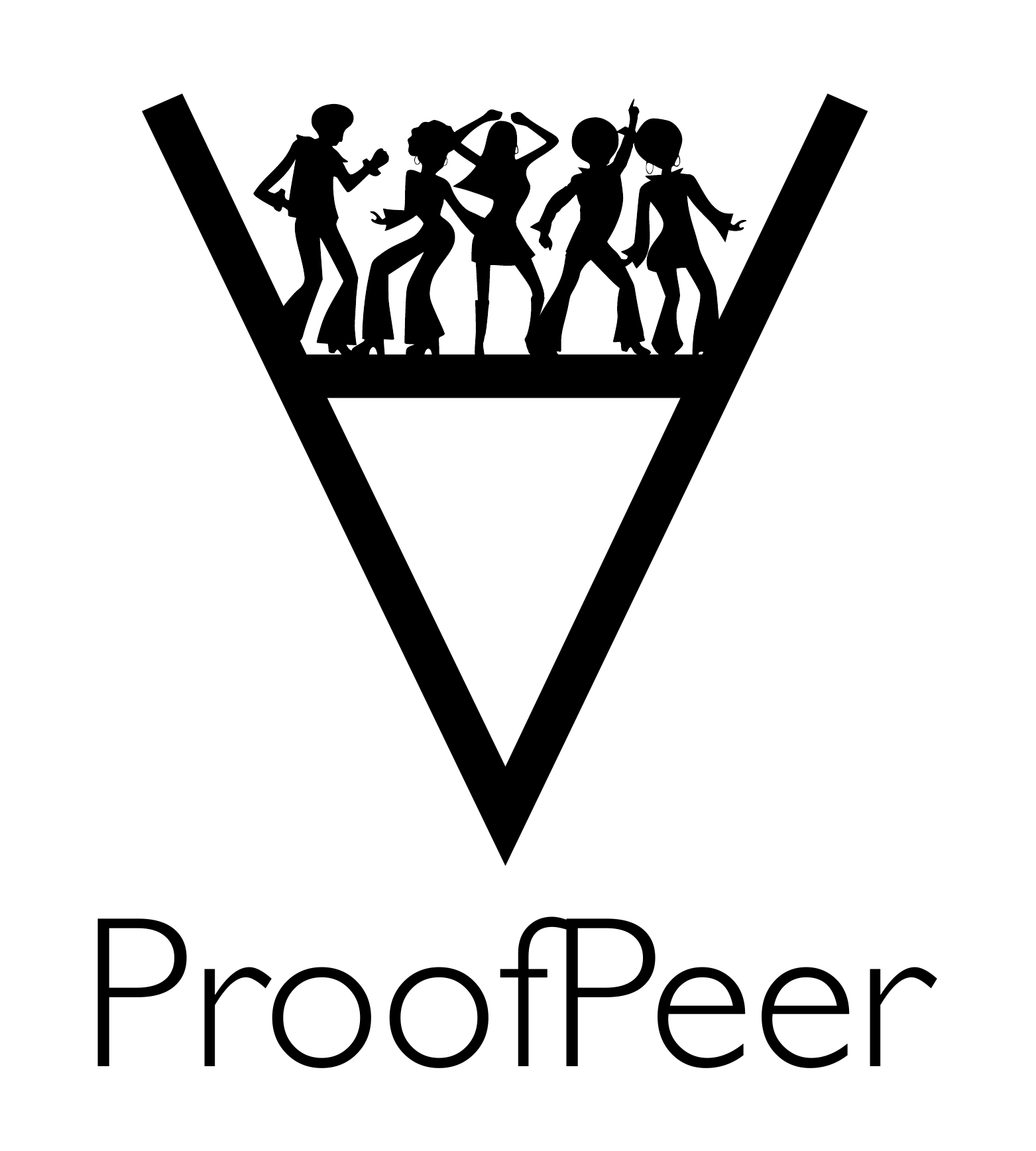}}
\end{center}

\vspace{1cm}

\begin{abstract}
We define the concept of \emph{collaborative theorem proving}
and outline our plan to make it a reality. We believe that a
successful implementation of collaborative theorem proving is a
necessary prerequisite for the formal verification of large
systems. 
\end{abstract}

\newpage
\section{The Challenge}

In today's computerised and scientific era, making our claims \emph{indefeasible} is more critical and more relevant than ever. We need to make indefeasible claims about the stability of mission critical hardware and software components~\cite{cipra_how_1995,gleick_bug_????}. We need to make indefeasible claims about technology used in medicine where mistakes literally \emph{kill}~\cite{_open-source_????}. And we are now entering an age where malware is transforming from a nuisance into a weapon of warfare~\cite{farwell_stuxnet_2011}.

\subsection*{Mechanised Theorem Proving} 
In response to the challenge of making indefeasible claims, the fields of \emph{automated} and \emph{interactive theorem proving} have emerged. \emph{Automated theorem proving} (ATP) is the mechanical checking of computerised proofs by mostly black box software components. 
\emph{Interactive theorem proving} (ITP) builds on this by allowing human insight to guide and coordinate ATP systems in almost arbitrary ways as they struggle in non-trivial domains.

\begin{figure}
\centering
\includegraphics[width=65mm]{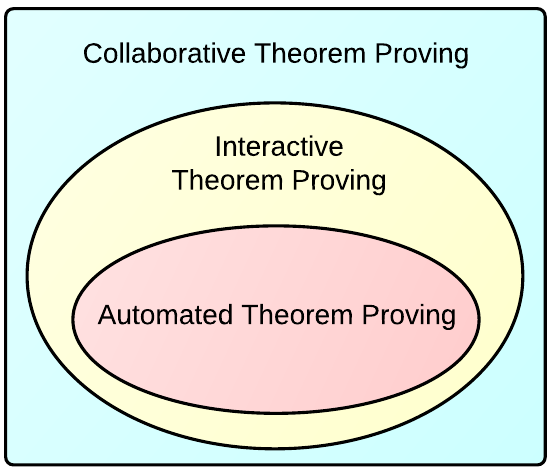}\\
\small Figure 1: Collaborative Theorem Proving
\end{figure}

ITP has had noteworthy successes. It has been used to prove the \emph{four colour theorem}~\cite{gonthier_formal_2008} in Coq~\cite{_coq_????}, the correctness of the \emph{seL4 microkernel}~\cite{klein_sel4:_2009,hutchison_challenges_2012} in Isabelle~\cite{_isabelle_????}, the correctness of a (large subset of a) C compiler in Coq~\cite{leroy_formal_2009}, and most recently, the proof of the \emph{Feit-Thompson odd order theorem}~\cite{gonthier_mathematical_2012} in Coq. Finally, the complete mechanisation of the formal proof of the Kepler conjecture in HOL~Light~\cite{_hol_????} is the ongoing goal of the \emph{Flyspeck project}~\cite{hales_revision_2010}. 

\begin{table}[t]
\centering
\begin{tabular}{l|ccc}
  & \textbf{Duration} & \textbf{Number of people} & \textbf{Lines of verification code} \\\hline
\textbf{Four colour theorem} & 2000 - 2005 & 1 & 60\,000 \\
\textbf{seL4 microkernel} & 2004 - 2009 & 17 & 200\,000 \\
\textbf{CompCert verified compiler} & 2005 - 2008 & 1 - 3 & 42\,000 \\
\textbf{Feit-Thompson theorem} & 2006 - 2012 & 15 & 170\,000 \\
\textbf{Flyspeck project} & 2003 - ? & 16 & 325\,000 
\end{tabular}
\caption{Major mechanised theorem proving efforts}
\label{bigver}
\end{table}

With these projects, we already have evidence that we can indefeasibly verify claims of both industrially relevant software \emph{and} deep mathematical problems, \emph{all using the same technology}. Table~\ref{bigver} lists the mentioned projects together with their duration and a rough estimate of their size in terms of lines of verification code. The scale here is impressive and might come as a surprise to those outside ITP circles. However, these are toy examples compared to serious problems such as the verification of a software system as large as, say, the Linux operating system kernel. Here, we would need to tackle a problem that is based on (as of Linux 3.2) about \emph{fifteen million lines of C code}, contributed by \emph{over 1300 developers} and \emph{over 200 companies}. 
The following seems obvious to us: \emph{Linux itself is the result of a large collaborative effort, therefore verifying it (if possible at all) will require an \textbf{even larger collaborative effort}. }

While we manage to develop software at scale, it is as yet unknown how to do verification at scale. 
We therefore propose to expand the capabilities of ITP by creating
\textbf{collaborative theorem proving} (Figure 1). We shall take the lessons and technologies of interactive theorem proving and integrate the lessons and technologies of the \emph{social web}~\cite{chi_social_2008}, in particular: 1) \emph{commons-based peer production}~\cite{benkler_coases_2002,benkler_commons-based_2006} and \emph{crowd-sourcing}~\cite{howe_rise_2006}, 
2) \emph{reputation systems}~\cite{resnick_reputation_2000} and 3) \emph{recommender systems}~\cite{ricci_introduction_2011, segaran_programming_2007}. 

\subsection*{The Web} 
One obvious step towards integrating theorem proving and web technology is to simply publish machine-checked proofs on the web. Prominent examples are the \emph{Mizar Mathematical Library}~\cite{_mizar_????-1} (MML) and the \emph{Archive of Formal Proofs}~\cite{_archive_????} (AFP). Both libraries are hosted at a central location and managed by a small fixed committee to which mechanised proofs are submitted. There is an inevitable maintenance burden~\cite{grabowski_revisions_2007}, as changes to the underlying theorem proving system might render previously valid proofs invalid.

An alternative approach is \emph{Logiweb}~\cite{grue_layers_2007}, which dispenses with the central authority and advocates a distributed model. Each user hosts their own Logiweb server, and while they may reference proofs hosted on other servers, the curation of the theory libraries themselves is down to the individual.


\subsection*{The Social Web} 
The next step beyond mere publication of mechanised proofs on the web is to add interactivity. To this end, the goal of the newest version of the \emph{Nuprl} ITP system~\cite{_nuprl_????}  is to be primarily accessible as a web service on top of which web user interfaces could be implemented. A similar approach is to equip an existing ITP system with a web interface~\cite{kaliszyk_web_2007, hutchison_web_2012, clide_cicm}.

We think a truly \emph{cohesive} system for \emph{collaborative theorem proving} (CTP) will be obtained by going further than merely plumbing together existing ITP systems and web software. We define collaborative theorem proving via
\begin{center}
CTP = The social machine of ITP.
\end{center}
The term \emph{social machine} was coined by Berners-Lee in 1999 to describe processes on the web where people do the creative work and computers play the role of administrators and assistants~\cite{berners-lee_weaving_2008}. This is exactly how we see collaborative theorem proving. 

We are interested in applying the following ideas from the social web to ITP:

\medskip
\noindent \textbf{1. Commons-based peer production and crowd-sourcing.} 
A prime example for \emph{commons-based peer production} (CBPP)~\cite{benkler_coases_2002,benkler_commons-based_2006} is \emph{Wikipedia}~\cite{_wikipedia_????}. It is regularly used by about half of the Internet population. The number of actual contributors is around 30\,000 per month.
Its success, and the success of the general wiki format, has inspired \emph{Wikiproofs}~\cite{_wikiproofs_????}, which in turn has been inspired by the 
\emph{Metamath}~\cite{_metamath_????} system, while ITP researchers have combined a wiki frontend with the Coq~\cite{kanade_large_2011} and Mizar~\cite{urban_wiki_2010} systems.

Regarding mathematics, a smaller, but nevertheless significant experiment is the \emph{Polymath project}, which uses an online forum open to anyone. The goal is to collaborate on finding new proofs of mathematical theorems. The first of a series of such projects was the density
 Hales-Jewett theorem, which was started by a post on the blog of Fields Medallist Timothy Gowers as an example of what he calls \emph{massively collaborative mathematics}~\cite{gowers_massively_2009}.

Another example of CBPP, mentioned already, is the Linux kernel. This project spawned off the version control system \emph{Git} which is the basis for another example of CBPP: \emph{Github}~\cite{_github_????}. Github is a site for hosting and sharing source code. It has around 3 million users sharing over 2.6 million public source code repositories. 

Closely related to CBPP is its commercial cousin \emph{crowd-sourcing}~\cite{howe_rise_2006}. Examples for crowd-sourcing are sites like \emph{Kaggle}~\cite{_kaggle_????}, which has attracted sixty thousand scientists who compete to produce the best solutions in machine learning problems, and \emph{Amazon mechanical turk}~\cite{_amazon_????-1} which connects over 260\,000 HITs (human  intelligence tasks) to 3 million workers. 

To apply CBPP and crowd-sourcing, it must be possible to decompose a task into many smaller tasks. There must then be a way to reintegrate the results of the smaller task into a result for the larger task. Mechanised theorem proving has unique advantages here. \emph{It can \textbf{guarantee} two things: that the smaller tasks completed correctly; and that the reintegration is valid}.

This potential for crowd-sourcing proof is also pursued by the DARPA project \emph{Crowd Sourced Formal Verification}\cite{darpacsfv} which started in 2011 and attacks the problem of formal verification of computer software by providing innovative user interfaces, for example in the form of iPad games. This project is different to what we are proposing in that we are focusing on extending ITP with the capabilities of the social web, while they are working on specific ways of making theorem proving accessible to laymen who might not even realise that their output is repurposed for theorem proving. It is imaginable that both approaches could be very fruitfully combined (see also Section 2 / Task 5).

\medskip
\noindent \textbf{2. Reputation systems.}
\emph{Stack Overflow}~\cite{_stack_????} is a Q\&A site for programmers, with around 1.5 million users, and a total of 4 million questions and 7.5 million answers. \emph{Math Overflow}~\cite{_math_????} is a Q\&A site for mathematicians, with around 23\,000 users and 40\,000 questions asked and answered. 

Both sites rely on commons-based peer production. To control for quality, they employ \emph{reputation systems}~\cite{resnick_reputation_2000}. Reputation is measured by \emph{reputation points}, which are usually earned when a user adds content, such as questions, answers, links to news and comments, that other users ``vote up''. In this way, they increase the content producer's reputation, who is then afforded more power at the site. For example, they might earn the ability to delete items submitted by other users.

We will use reputation to instill \emph{confidence}. Confidence is already guaranteed for mechanically verified theorems, but collaborative theorem proving is broader than this. Confidence is also needed in the process of library curation, burdening a centralised panel of expert curators (as in the case of the MML or the AFP). We intend to use reputation to replace this centralised authority by a dynamic committee consisting of peers with high reputation. We will also use reputation to deal with another important confidence issue. Peers are more likely to invest time on small problems within a large verification project if they are confident that their work will pay off in solving the larger problem. This means they must have confidence in the original analysis of the problem and its proposed decomposition into subproblems. They can gain this confidence if those proposing the decomposition have high reputation. 

Using reputation to instill confidence in the decomposition mirrors the reality of large verification projects like Flyspeck and the mechanisation of the Feit-Thompson theorem, the decompositions of which are managed by Hales and Gonthier respectively.

\medskip
\noindent \textbf{3. Recommender systems.} Hales estimates\footnote{personal communication by email} that almost $50\%$ of his time formalising the Jordan curve theorem and working on Flyspeck was spent looking up theorems in the HOL~Light library, which suggests we need much better search tools in ITP. Contemporary approaches are all pattern-based: given a search pattern, look up the corresponding matches in the library, and possibly take subpatterns into account. This approach is pursued, for example, with Whelp~\cite{asperti_content_2006}, MathWebSearch~\cite{kohlhase_mathwebsearch_2012} and MIaS~\cite{sojka_indexing_2011}. 

An alternative is to use \emph{recommender systems}~\cite{ricci_introduction_2011, segaran_programming_2007}. These are software tools and techniques which provide suggestions for items of use to a user. For theorem proving, these items would typically be theorems, but they could just as well be \emph{peers} of interest. Finding these theorems and peers will use the two main techniques of recommender systems (often combined in hybrid approaches): \emph{collaborative filtering} and \emph{content-based filtering}. 

Recommender systems typically include a ranking component which can be coupled with reputation systems. They are widely deployed in the industry, e.g. collaborative filtering is used by \emph{Netflix}~\cite{bell_lessons_2007} for making movie recommendations.  

Preliminary work by Fleuriot on modifying and applying the Slope One item-based collaborative filtering algorithm~\cite{lemire_slope_2005} to the Jordan Curve Theorem corpus in HOL~Light indicates that it can offer relevant theorem recommendations when tested against previously unseen (but already mechanised) results.


Content-based recommender systems are already an important part of ITP. Tools like Sledgehammer~\cite{blanchette_extending_2011} use relevance filtering~\cite{meng_lightweight_2009} and premise selection techniques~\cite{kuhlwein_overview_2012} to scan the context of an interactive proof for useful theorems. Once identified, these theorems are passed to ATP systems which are often able to prove the goal. While the success rate is highly dependent on the quality of the context information, astonishing $40\%$ success rates have been reported~\cite{kaliszyk_learning-assisted_2012}. Other advanced machine learning techniques not only recommend theorems, but \emph{proof patterns}~\cite{heras_statistical_2013}. 

\section{Implementing Collaborative Theorem Proving}

We think that:
\begin{quote}\it
\textbf{Collaborative theorem proving} can tackle verification tasks of unprecedented magnitude.
\end{quote}
Under this assumption, the obvious question is: how do we make
collaborative theorem proving a reality? In the remainder of this
paper, we will outline our plan for doing so. In particular:

\begin{itemize} 
\item We will prove that the technical requirements of CTP can be met,
  by defining mechanisms for collaborative theorem proving: these include 1) machinery for allowing commons-based peer production and crowd-sourcing,
2) a reputation system that applies to peers and the artefacts created by them, and 3) a hybrid recommender system based on \emph{both} collaborative and content-based filtering which supports the theorem proving activities of peers and also feeds into automation. 
\item We will grow a community called \textbf{ProofPeer} which uses an experimental research prototype which implements these mechanisms for collaborative theorem proving. This prototype will be called ProofPeer, and will be hosted as a central hub at  \mbox{\url{http://proofpeer.net}}. It will be accessible by any modern web browser and will not require any additional installation on the user's behalf. This keeps the barriers to entry low, so as not to discourage participants. 
\item We will show that the community can create a mechanically verified library of mathematical theorems. We will measure the speed at which this library  is being created, the quality of the library and the size of the library, and compare these measurements to
what we know about the growth of libraries for current popular proof
assistants. Most importantly, we will measure the extent to which the
mechanisms for CTP are being used to create this library, and try to
determine the impact that these mechanisms have on growth speed,
quality and size. Note that at this stage, it is not obvious how to
define ``quality''. Coming up with an appropriate definition is part
of the challenge.
\end{itemize}

\subsection*{Growing a Community}

A key challenge of our plan is that we depend on having a community of
peers that we can experiment on. Such a community will allow us to
continuously gather data and combine theory and experiment in an
iterative feedback loop. We will solve this problem by 
introducing students to ProofPeer as part of their course and project
work and by making the system available to members of
the research community. At some point, the system will be open to the public, and we will reach out to the wider audience to recruit peers. We must be able to cope with any potential influx of new users, so one of our goals is to have a system which can scale to arbitrary numbers of peers.
To provide formalisation goals, we will setup many mathematical theorems as conjectures (see Task 1) in ProofPeer and award reputation points for proofs of them. This list will be seeded from Wiedijk's list of theorems~\cite{_100_????} which contains theorems of wildly varying difficulty, ranging from basic theorems like the triangle inequality to celebrated theorems of mathematics like Fermat's last theorem. We will also add open problems like the Goldbach conjecture and problems that may be added in the future to the Polymath project.

We want to be flexible in our exploration of the design space of
collaborative theorem proving. In particular we want to deal with the
higher-level implications of collaboration and scale, and not so much
with the lowest-level ones. We think we can best do this by creating a
new theorem proving system which can run on top of a proven cloud
computing platform like Google App Engine~\cite{_google_????}. This
platform already provides many important abstractions like a scalable
data store and scalable session management.  We also want a
\emph{quick start}; therefore, we strive to make our system compatible with
HOL~Light. This is a good choice of system for two reasons: firstly,
it contains one of the largest libraries of mechanically verified
mathematical theorems; secondly, it has a very simple and elegant
implementation which can be feasibly incorporated with ProofPeer. 

The following subsections break down our work programme in more detail.

\subsection*{Task 1: CBPP / crowd-sourcing}
The main problem to solve for effective CBPP / crowd-sourcing is how to manage the decomposition of a task and the reintegration of the resulting smaller tasks. To this end, we will study mechanisms for \emph{version control}, \emph{continuous integration}, \emph{conjectures / gaps} and \emph{access control}.

We will define and implement a mechanism for \textbf{version control}. We have already experimented with such a mechanism and have partially implemented it on top of Google App Engine as a versioned file system based on 3-way-merge and the diff3~\cite{khanna_formal_2007} merge algorithm for text files. We will need to determine how best to specialise such merging for theory files (written in ProofScript, see Task 4). 

Closely connected to version control is the issue of \textbf{continuous integration}. Formal theories depend on each other, so a change in one theory may imply that depending theories do not mechanically verify anymore, or are subject to more subtle changes. We will be looking at mechanisms to reduce costly and time consuming rechecking of theories, exploit the potential for parallel and asynchronous processing~\cite{wenzel_asynchronous_2012}, and also explore how far related work in the area of algebraic specifications~\cite{hutter_management_????} is applicable to our specific case.

A common phenomenon of formal verification is the presence of \textbf{conjectures} and more generally \textbf{gaps} in intermediate stages of the verification~\cite{dixon_proof-centric_2006,wiedijk_formal_2004}. We will study how to explicitly support such gaps in ProofPeer (this is interdependent with Task 4), and how to advertise the existence of such gaps to peers who might be willing to close them.  

Finally, we will be studying \textbf{access control} mechanisms. Peers might prefer to keep certain formal verification projects private or only accessible by a certain group of peers. Access control is particularly important for industrial verification projects where intellectual property is an issue. We will connect access control to the reputation system (Task 2) so that peers with high reputation can curate the library effectively.

\subsection*{Task 2: reputation system}
We will define and implement a reputation system for peers based on points, which should track the value of a user's contributions. A contribution can be, for example, a verified theorem, a definition, a conjecture, or the closing of a gap (see Task 1). It might also be a more informal artefact such as a \emph{comment} about another contribution, which itself might just be another comment. 

The value of a contribution could be influenced by where, how often and by whom it is used. It can also be influenced by other peers explicitly voting on that value, e.g. either by ``liking" or ``disliking" a contribution.  We will experiment with a special form of voting which consists of setting up a \textbf{bounty} for proving a conjecture (first).

There are links to the issue of version control from Task 1: how do we transfer direct votes from peers between different versions? Assume somebody made a change to the statement of a theorem. Will that negate all previous votes for this theorem? Or if a bounty has been set up for a conjecture, should that conjecture be allowed to change?

The goal of our reputation system is threefold: 1) to enable us to carve out a dynamic group of peers who have high reputation and who are granted special (access control) powers for library curation; 2) to motivate  peers; 3) to serve as input to our recommender system (see Task 3).

\subsection*{Task 3: recommender system}
There are two separate motivations for the use of recommender systems in ProofPeer. 

\medskip
\noindent 1) To allow a peer to explore topics of interest to them. These might be theorems, but they might also be other peers. To characterise the topics, we will add \textbf{social tagging}, and will then experiment with a recommender system that uses both the tags and the reputation points as input for suggesting interesting topics to peers. 

\medskip
\noindent 2) To support a peer during interactive theorem proving. We will define and implement a \textbf{context aware recommender system} that will suggest theorems and proof patterns that might be useful in a given theorem proving situation based on implicit and explicit observations. Our first iteration for such a recommender system will be content-based. We will then try to combine this content-based approach with the recommender system from 1). E.g., we might restrict the content we are using to that produced by peers with high reputation, or to content with related tags. 

\subsection*{Task 4: ProofScript}
We will define a language called \textbf{ProofScript} that will be the
primary interface between peers and the collaborative theorem proving
system. ProofScript will serve both as a \textbf{programming language}
(similar in function to Standard ML in HOL-based systems) and as a
\textbf{proof language} (similar to the \emph{Isabelle/Isar}
language). An important design constraint is that we need to make
conversion from HOL~light to ProofScript possible.

Starting from these initial design constraints we will codevelop ProofScript with the other features of ProofPeer. An example of why this is necessary lies in the fact that ProofScript should support \textbf{namespaces} which align with the hierarchal structure of our versioned filesystem. Another example is that ProofScript should have language constructs for stating \textbf{conjectures} and marking \textbf{gaps} within the formalisation which can be tracked by ProofPeer.


We will implement an interpreter for ProofScript which is deployed both on the server (i.e., on Google App Engine) and the client. We will try to shift most of the computing to the distributed peer machines by allowing proof scripts to run not only on the server, but also on the client. This means devising a strategy for how server and client side work together. Ideally, the client will be responsible for \emph{finding} proofs. \emph{Checking} those proofs is the duty of the server.

\subsection*{Task 5: user interface}
We will have the entire user interface running in a web browser. Our
starting point will be the Clide~\cite{clide_cicm} system which builds
on the PIDE~\cite{wenzel_isabelle/jedit_2012} approach to proof
interaction. We will also investigate integration of the proof editor / viewer with our hiproof visualisation~\cite{obua_capturing_2013}. Furthermore, we must provide appropriate user interface abstractions for dealing with all features of ProofPeer like versioning, and we will need to seek straightforward and intuitive ways for users to interact with the features developed in Tasks 1--4. These features will be exposed as part of an application programming interface (API), which will be the sole means by which the user interface will communicate with ProofPeer. This means other researchers can use the API to connect their own applications and interfaces to ProofPeer in a seamless way. For example, innovative user interfaces like Xylem~\cite{darpacsfv} for crowd-sourced proof via gaming could be integrated with ProofPeer so that certain proof obligations arising within ProofPeer are not handled entirely within the system, but are outsourced to Xylem instead, and then reintegrated with ProofPeer via proof certificates.     

\section{Conclusion}
This position paper is based on our EPSRC grant proposal \emph{EP/L011794/1}. The grant
started in March 2014 and will allow us to work on making
collaborative theorem proving a reality. We believe that one of the
main applications of CTP will be in specifying and verifying large
systems. And because growing a community
is such an important aspect of CTP we believe that it is important to
get in touch with possible beneficiaries of CTP such as systems
designers as soon as possible. 

\medskip
\noindent\textbf{Acknowledgements.} Many thanks to everyone who commented on
(earlier drafts) of our grant proposal and therefore indirectly also
on this paper, in particular: Alan Bundy, Brian Campbell, Ursula Martin, James McKinna, Alison
Pease, and Norbert Schirmer. 

\bibliographystyle{eptcs}
\bibliography{proofpeer}
\end{document}